\documentclass[sigconf]{acmart}
\usepackage{afterpage}

\AtBeginDocument{%
  \providecommand\BibTeX{{%
    \normalfont B\kern-0.5em{\scshape i\kern-0.25em b}\kern-0.8em\TeX}}}

\begin{document}

\title{CADC: Encoding User-Item Interactions for Compressing Recommendation Model Training Data}

\author{Hossein Entezari Zarch}
\email{entezari@usc.edu}
\affiliation{%
  \institution{University of Southern California}
  \city{Los Angeles}
  \state{California}
  \country{USA}
}

\author{Abdulla Alshabanah}
\email{aalshaba@usc.edu}
\affiliation{
  \institution{University of Southern California}
  \city{Los Angeles}
  \state{California}
  \country{USA}
}

\author{Chaoyi Jiang}
\email{chaoyij@usc.edu}
\affiliation{
  \institution{University of Southern California}
  \city{Los Angeles}
  \state{California}
  \country{USA}
}

\author{Murali Annavaram}
\email{annavara@usc.edu}
\affiliation{
  \institution{University of Southern California}
  \city{Los Angeles}
  \state{California}
  \country{USA}
}

\renewcommand{\shortauthors}{Entezari Zarch, et al.}

\begin{abstract}
  Deep learning recommendation models (DLRMs) are at the heart of the current e-commerce industry. However, the amount of training data used to train these large models is growing exponentially, leading to substantial training hurdles. The training dataset contains two primary types of information: content-based information (features of users and items) and collaborative information (interactions between users and items). One approach to reduce the training dataset is to remove user-item interactions. But that significantly diminishes collaborative information, which is crucial for maintaining accuracy due to its inclusion of interaction histories. This loss profoundly impacts DLRM performance.
 This paper makes an important observation that if one can capture the user-item interaction history to enrich the user and item embeddings, then the interaction history can be compressed without losing model accuracy. Thus, this work, Collaborative Aware Data Compression (CADC), takes a two-step approach to training dataset compression. In the first step, we use matrix factorization of the user-item interaction matrix to create a novel embedding representation for both the users and items. Once the user and item embeddings are enriched by the interaction history information the approach then applies uniform random sampling of the training dataset to drastically reduce the training dataset size while minimizing model accuracy drop. The source code of CADC is available at \href{https://anonymous.4open.science/r/DSS-RM-8C1D/README.md}{https://anonymous.4open.science/r/DSS-RM-8C1D/README.md}.
\end{abstract}




\maketitle

\section{Introduction}
Deep learning recommendation models (DLRM) play a pivotal role in enhancing user experience, by suggesting new and pertinent content, across numerous online platforms. Companies like Meta, Google, Microsoft, Netflix, and Alibaba employ these sophisticated models for a range of services, including personalizing and ranking Instagram stories \cite{medvedev2019powered}, video suggestions on YouTube \cite{covington2016deep}, mobile app recommendations on Google Play \cite{cheng2016wide}, personalized news and entertainment options \cite{elkahky2015multi, steck2021deep}, and tailored product recommendations \cite{zhou2019deep}. Additionally, tasks such as Newsfeed Ranking and Search are also built upon DNNs \cite{gupta2020deeprecsys, gupta2020architectural, naumov2019deep, song2020towards}, further exemplifying the critical role these models play in content discovery and user engagement.

The essential role of DLRMs in generating revenue for many internet companies has led to their marked increase in complexity and size. 
From 2017 to 2021, the  Meta's DLRM model size escalated 16-fold, requiring terabytes of model weights~\cite{mudigere2021high}. Additionally, the need for memory bandwidth to manage these models increased almost 30-fold\cite{sethi2022recshard}. This growth translates to recommendation models consuming over 50\% of training and 80\% of AI inference cycles\cite{acun2021understanding, gupta2020architectural, naumov2020deep, lui2021understanding, zhao2020distributed}. The essential role of DLRMs in generating revenue for many internet companies has led to their marked increase in complexity and size. From 2017 to 2021, the Meta’s DLRM model size escalated 16-fold, requiring terabytes of model weights. Additionally, the need for memory bandwidth to manage these models increased almost 30-fold. This growth translates to recommendation models consuming over 50\% of training and 80\% of AI inference cycles. A similar phenomenon in transformer-based models, including large language models, and vision transformers, has been comprehensively tackled by various techniques targeting parameter- and memory-efficient training  \cite{DBLP:conf/iclr/HuSWALWWC22,DBLP:journals/corr/abs-2308-03303,azizi2024lamda,DBLP:journals/corr/abs-2310-11454} and inference acceleration \cite{DBLP:journals/corr/abs-2210-17323,sadeghi2024peano,DBLP:conf/nips/KwonKMHKG22,DBLP:conf/eccv/YuanXCWS22}. While these complexities are increasingly managed in transformer-based models, DLRMs have not seen the same level of in-depth analysis. As DLRMs expand, so too does the size of their training datasets, which are comprised primarily of user-item interactions. Thus, the system infrastructure, such as GPU and CPU count, and total system memory, to support these models has grown by up to 2.9 times in just a few years \cite{wu2022sustainable, mudigere2022software}. 

One aproach to reduce the training costs is to reduce the training dataset size. There are also orthogonal approaches to reduce the model size  but this work focuses on reducing the training dataset size. 
DLRMs derive value from two primary types of information within large datasets: content-based information (features of users and items) and collaborative information (interactions between users and items). Removing interactions from large datasets significantly diminishes collaborative information, which is crucial for maintaining accuracy due to its inclusion of interaction histories. This loss profoundly impacts DLRM performance.

To address this challenge, we propose Collaborative Aware Data Compression (CADC), a strategy that harnesses Matrix Factorization (MF), a computationally efficient model, to capture and compress the entire collaborative spectrum of a dataset into a set of user and item embeddings. This method ensures that the essential collaborative information is preserved, even when the dataset undergoes substantial reduction. By using these pre-trained embeddings in DLRMs, the models are less sensitive to collaborative interaction data loss when the dataset is filtered. We test our approach on Movielens 1M and 10M, and Epinions, showing that it can keep models accurate even after many of the user-item interactions have been removed.


\section{Collaborative Aware Data Compression}
\label{sec:cadc}
Let \( \mathcal{D}_{\text{full}} \) denote the entire training dataset, consisting of users \( \mathcal{U} \) and items \( \mathcal{V} \). Our objective is to train a base Two Tower Neural Networks (TTNN) model, represented by \( \mathcal{M}_{\text{TTNN}} \), on \( \mathcal{D}_{\text{full}} \). Due to the large size of \( \mathcal{D}_{\text{full}} \), direct training is impractical. To address this, we create \( \mathcal{D}_{\text{sel}} \), a subset in which interactions are randomly selected from \( \mathcal{D}_{\text{full}} \) to ensure that it includes a representative sample of the original \( \mathcal{U} \) and \( \mathcal{V} \) \. This selection process is designed to preserve the statistical properties and data distribution of \( \mathcal{D}_{\text{full}} \), thereby reducing the computational demands of training \( \mathcal{M}_{\text{TTNN}} \) without compromising the integrity of the dataset's inherent structure.

We build on the observation that collaborative information present in user-item interactions must be captured for ensuring model accuracy  when reducing dataset size. To address this, we introduce the CADC technique. This method involves training a compact collaborative filtering model, specifically Matrix Factorization (MF), on the collaborative information residing in \( \mathcal{D}_{\text{full}} \) to generate pre-trained embeddings for \( \mathcal{U} \) and \( \mathcal{V} \) \ based on their complete interaction profiles. These embeddings are then integrated into the \( \mathcal{M}_{\text{TTNN}} \). Incorporating these pre-trained weights, which encapsulate the entire collaborative information from \( \mathcal{D}_{\text{full}} \), allows \( \mathcal{M}_{\text{TTNN}} \) to access comprehensive interaction data while only being trained on a significantly smaller subset, \( \mathcal{D}_{\text{sel}} \). This strategic integration dramatically mitigates the adverse effects of training data filtering and preserves model accuracy by maintaining vital collaborative information inside the DLRM. The ensuing sections detail the CADC methodology.
\subsection{Pre-training Embedding Vectors}
To encapsulate the entire collaborative information within \( \mathcal{D}_{\text{full}} \) into a set of embedding vectors, we employ MF, a well-regarded and computationally efficient method in collaborative filtering. Given the massive size of \( \mathcal{D}_{\text{full}} \), the training methodology must not only be computationally efficient but also capable of capturing the dynamics of user-item interactions with high fidelity. This method efficiently captures the dynamics of user-item interactions by reducing the high-dimensional interaction space into a lower-dimensional, continuous feature space.

In MF, we construct two separate embedding tables: one for users and another for items. Each user and item is represented as a vector, \( \mathbf{u}_i \) and \( \mathbf{v}_j \), in this latent feature space. To enhance the model's ability to capture individual preferences and item qualities, we incorporate a bias term for each user and item into their respective vectors. The last element of each vector, \( u_{i, \text{bias}} \) and \( v_{j, \text{bias}} \), serves as this bias term. The interaction between user $i$ and item $j$ is formulated as follows:
\[
\hat{y}_{\text{MF}}(i, j) = \sigma((\mathbf{u}_i^\prime \cdot \mathbf{v}_j^\prime) + u_{i, \text{bias}} + v_{j, \text{bias}} + b)
\]

where \( \mathbf{u}_i^\prime \) and \( \mathbf{v}_j^\prime \) represent the vectors \( \mathbf{u}_i \) and \( \mathbf{v}_j \) excluding their respective bias terms \( u_{i, \text{bias}} \) and \( v_{j, \text{bias}} \). \( b \) represent the global bias term. \( \sigma \) is the sigmoid function.

To optimize these embeddings, we employ binary cross-entropy loss. Due to the implicit feedback nature of the datasets and scarcity of positive interactions, we implement negative sampling to balance the labels' distribution. Specifically, we generate \( \mathcal{D}_{\text{neg}} \), a subset of negative interactions, to counteract the sparsity of the data where most labels are implicitly zero. The loss formulation, which incorporates both sets of interactions, is as follows:
\[
\mathcal{L}_\text{MF} = -\sum_{(i, j) \in \mathcal{D}_{\text{full}}} \log(\hat{y}_{\text{MF}}(i, j)) - \sum_{(i, j) \in \mathcal{D}_{\text{neg}}} \log(1-\hat{y}_{\text{MF}}(i, j))
\]

\subsection{Integrating Enhanced Embedding Vectors into DLRM}
In our approach, we employ the TTNN as a variant of the DLRM, which processes user and item features through distinct pathways. Each pathway comprises a Multi-Layer Perceptron (MLP): the user tower, \( \mathcal{T}_{\text{user}} \), and the item tower, \( \mathcal{T}_{\text{item}} \). Specifically, \( \mathcal{T}_{\text{user}} \) processes a concatenated vector of the corresponding user identifier and its features, \( \mathcal{F}_{\text{user}, i} = [\text{id}_{\text{user}, i}; \text{features}_{\text{user}, i}] \), while \( \mathcal{T}_{\text{item}} \) handles a similar vector for items, \( \mathcal{F}_{\text{item}, j} = [\text{id}_{\text{item}, j}; \text{features}_{\text{item}, j}] \).

The interaction between a user \( i \) and an item \( j \) is modeled by the dot product of the embedding vectors output by each tower, followed by a sigmoid transformation to compute the prediction score. This is formally expressed as:
\[
\hat{y}_{\text{TTNN}}(i, j) = \sigma(\mathcal{T}_{\text{user}}(\mathcal{F}_{\text{user}, i})^T \cdot \mathcal{T}_{\text{item}}(\mathcal{F}_{\text{item}, j}))
\]
where \( \sigma \) denotes the sigmoid function.

The enriched embeddings obtained from our pre-training step using MF, \( \mathbf{u}_i \) and \( \mathbf{v}_j \), are used to initialize and freeze the identifiers \( \text{id}_{\text{user}} \) and \( \text{id}_{\text{item}} \) within \( \mathcal{F}_{\text{user,i}} \) and \( \mathcal{F}_{\text{item,j}} \) with these vectors. Consequently, the TTNN's identifier embedding tables become non-trainable, and \( \mathcal{T}_{\text{user}} \) and \( \mathcal{T}_{\text{item}} \) now process the inputs \( \mathcal{F}_{\text{user}, i} = [\mathbf{u}_i; \text{features}_{\text{user}, i}] \) and \( \mathcal{F}_{\text{item}, j} = [\mathbf{v}_j; \text{features}_{\text{item}, j}] \) respectively. \( \mathcal{M}_{\text{TTNN}} \) is then trained on \( \mathcal{D}_{\text{sel}} \), leveraging the comprehensive interaction dynamics encapsulated within \( \mathcal{D}_{\text{full}} \). This sophisticated integration not only leverages the depth of neural networks but also harnesses the breadth of collaborative filtering, ensuring a robust and accurate prediction mechanism.

Findings detailed in Section \ref{sec:sense} reveal that using non-trainable embeddings for user and item identifiers in \( \mathcal{M}_{\text{TTNN}} \) is the most effective approach, achieving the highest model accuracy among evaluated methods. This method, by freezing the embeddings, simplifies the training process as it does not require updating the large user and item ID embedding tables—typically the most substantial component in a DLRM. 
When the TTNN is trained on the reduced dataset \( \mathcal{D}_{\text{sel}} \), it effectively utilizes the comprehensive interaction dynamics originally captured within \( \mathcal{D}_{\text{full}} \). This approach not only exploits the depth of neural networks but also incorporates the extensive capabilities of collaborative filtering, offering a robust and precise mechanism adept at managing the challenges posed by large-scale industry datasets.

\textbf{Compression:} In the last step of our approach we randomly sample a subset of the user-item interactions to create a compressed training dataset. Since our pre-training captures the interaction history efficiently we do not need to explore any complex compression schemes, as our results show next.   

\section{Experimental Setup}
Our experiments were designed to assess the efficacy of the CADC across three distinct datasets, each with unique characteristics: MovieLens 1M\footnote{The MovieLens 1M dataset, available at \url{https://grouplens.org/datasets/movielens/1m/}}, MovieLens 10M\footnote{The MovieLens 10M dataset, available at \url{https://grouplens.org/datasets/movielens/10m/}}, and Epinions\footnote{The Epinions dataset, available at \url{https://alchemy.cs.washington.edu/data/epinions/}}. For evaluation, the last two interactions of each user were reserved for validation and testing. Each dataset was subjected to training over 100 epochs on $10\%$ of its interaction data using a TTNN architecture, with embedding sizes set to 96. Before training the TTNN, an MF model was trained on all interactions within each dataset for 100 epochs, with an embedding size of 95. This size aligns with the TTNN's effective size when incorporating additional bias terms for users and items. The optimization of MF employed an alternating scheme using the Adam optimizer: item embeddings were fixed while updating user embeddings, and vice versa.

Performance was evaluated using Hit Rate at 10 \textbf{(HR@10)} and Normalized Discounted Cumulative Gain at 10 \textbf{(NDCG@10)}, metrics that assess accuracy and ranking quality. Additionally, the training time for each dataset was recorded in seconds to evaluate the time efficiency of the method. In all scenarios utilizing CADC, a small model was first trained on the entire dataset, after which the embeddings were transferred to the TTNN and frozen, as detailed in Section \ref{sec:cadc}.

\vspace{-2mm}
\subsection{Baselines}
To evaluate the effectiveness of the CADC, it was benchmarked against various methods:

\begin{itemize}
    \item \textit{Random}: This baseline trains the TTNN on the filtered dataset without any sophisticated data compression or embedding optimization techniques, serving as a naive control.
    \item \textit{Long-Tail Item Recommendation Techniques (Over-Sampling, Under-Sampling, LogQ)} \cite{zhang2021model}: These methods are incorporated as baselines to address challenges posed by data filtering, which often exacerbates the long-tail problem in recommendation systems.
    \item \textit{CADC-MLP} \cite{he2017neural}: This variant employs an MLP as the interaction function between user and item embeddings, replacing the traditional dot-product approach used in MF. It offers a more complex interaction model, making it computationally more intensive than traditional MF.
\end{itemize}

\section{Results}
\begin{table*}[!t]
    \centering
    \begin{tabular}{c|cc|c|cc|c|cc|c}
        \hline
        Method & \multicolumn{3}{c|}{MovieLens 1M} & \multicolumn{3}{c|}{MovieLens 10M} & \multicolumn{3}{c}{Epinions} \\
        \cline{2-10}
        & HR@10 & NDCG@10 & Time & HR@10 & NDCG@10 & Time & HR@10 & NDCG@10 & Time \\
        \hline
        Random & 3.76 (45.7\%) & 1.71 (49.6\%) & 142 & 5.25 (42.9\%) & 2.62 (41.6\%) & 848 & 1.53 (60.3\%) & 0.70 (63.7\%) & 15 \\
        Over-Sampling & 0.70 (89.9\%) & 0.32 (90.6\%) & 142 & 0.09 (99.0\%) & 0.04 (99.1\%) & 848 & 0.63 (83.6\%) & 0.26 (86.5\%)& 15 \\
        Under-Sampling & 1.84 (73.4\%) & 0.84 (75.2\%) & 142 & 1.74 (81.1\%) & 0.80 (82.2\%) & 848 & 0.98 (74.5\%) & 0.42 (78.2\%) & 15 \\
        LogQ & 4.21 (39.2\%) & 1.99 (41.3\%) & 142 & 5.63 (38.7\%) & 2.73 (39.2\%) & 848 & 1.69 (56.1\%) & 0.90 (53.4\%) & 15 \\
        CADC-MLP & 6.23 (10.0\%) & 2.95 (13.0\%) & 302+136 & 8.48 (7.7\%) & 4.17 (7.1\%) & 3007+753 & 3.55 (7.8\%) & 1.82 (5.7\%) & 44+14 \\
        CADC & \textbf{6.57 (5.1\%)} & \textbf{3.28 (3.2\%)} & 18+136 & \textbf{8.54 (7.1\%)} & \textbf{4.18 (6.9\%)} & 184+755 & \textbf{3.79 (1.6\%)} & \textbf{1.89 (2.1\%)} & 12+14 \\
        \hline
        GS & 6.92 & 3.39 & 1464 & 9.19 & 4.49 & 8299 & 3.85 & 1.93 & 142\\
        \hline
    \end{tabular}
    \caption{Performance comparison of different data handling strategies using the MovieLens 1M, MovieLens 10M, and Epinions datasets. Metrics evaluated include HR@10 and NDCG@10 for recommendation accuracy and computational time in seconds. The percentages in parentheses indicate the performance reduction from the Gold Standard (GS).}
    \vspace{-8mm}
    \label{tab:mresults}
\end{table*}

Table \ref{tab:mresults} presents the performance analysis of the recommendation systems utilizing the CADC method compared to other approaches across three datasets: MovieLens 1M, MovieLens 10M, and Epinions. Performance metrics include HR@10, NDCG@10, and computational time (seconds). The parentheses values indicate the percentage performance degradation relative to the Gold Standard (GS).

CADC achieved notable success in maintaining high recommendation quality while training on only $10\%$ of the data, significantly reducing the training time compared to models trained on the entire dataset. Specifically, for MovieLens 1M, CADC showed superior performance, achieving an HR@10 of 6.57 and an NDCG@10 of 3.28, with only a $5.1\%$ and $3.2\%$ degradation in performance compared to the GS, respectively. These results were obtained with significantly reduced computational time (154.6 seconds compared to 1464.4 seconds for the GS).

In the larger MovieLens 10M dataset, CADC continued to outperform other methods with an HR@10 of 8.54 and an NDCG@10 of 4.18, marking a $7.1\%$ and $6.9\%$ performance decline relative to the GS. The computational time was notably lower (939.6 seconds) compared to the GS, which required 8298.8 seconds. For the Epinions dataset, CADC demonstrated the highest performance improvements, with an HR@10 of 3.79 and an NDCG@10 of 1.89, indicating minimal performance degradation of $1.6\%$ and $2.1\%$ respectively compared to the GS, achieved in significantly lesser time (26.4 seconds).

The Random baseline method exhibited substantial performance declines across all datasets, with the most significant reductions observed in the Epinions dataset, where HR@10 was only 1.53 and NDCG@10 was 0.70. The CADC-MLP variant, although performing better than the Random baseline, was still consistently outperformed by the CADC method in terms of both ranking quality and time efficiency. The experimented sampling methods, including Over-Sampling and Under-Sampling, failed to yield promising results. The LogQ method achieved results that were marginally better than those of the naive Random approach but remained substantially inferior to the GS.

\section{Sensitivity Analysis}
\label{sec:sense}
This section examines the impact of various factors on the performance of the CADC method, specifically focusing on the data filtering ratio and different embedding integration techniques. All experiments were conducted using the MovieLens-1M dataset.

\begin{itemize}
    \item \textbf{Data Filtering Ratio:} Defined as the ratio of the number of interactions in $\mathcal{D}_{\text{full}}$ to those in $\mathcal{D}_{\text{sel}}$. For instance, a data filtering ratio of 50 means that the DLRM is trained on 2\% of the complete dataset. Figures illustrate that as the data filtering ratio increases, indicating more substantial data reduction, the decline in ranking performance becomes progressively less pronounced. This demonstrates CADC's capability to maintain model accuracy effectively, even with significantly reduced datasets.

    \item \textbf{Embedding Integration Techniques:} Analysis of different integration techniques within the CADC framework reveals varied impacts on performance:
    \begin{itemize}
        \item \textit{Hybrid}: Employs a combination of pre-trained and trainable elements where two-thirds of the id embedding elements are derived from MF and the remaining third are regular trainable parameters.
        \item \textit{Init}: Initializes the id embeddings entirely with pre-trained vectors, which remain updatable during training, offering flexibility in adaptation.
        \item \textit{Init-Frz}: Currently used in CADC, this method involves initializing the id embeddings with pre-trained vectors and subsequently freezing them, enhancing stability.
        \item \textit{Linear}: Similar to \textit{Hybrid}, but incorporates a trainable linear layer that processes the pre-trained embeddings before their integration.
        \item \textit{MLP}: Builds on the \textit{Linear} approach by applying an MLP to the pre-trained embeddings, introducing a higher level of model complexity.
    \end{itemize}
\end{itemize}

\begin{figure}[htbp]
\centering
\includegraphics[width=\linewidth]{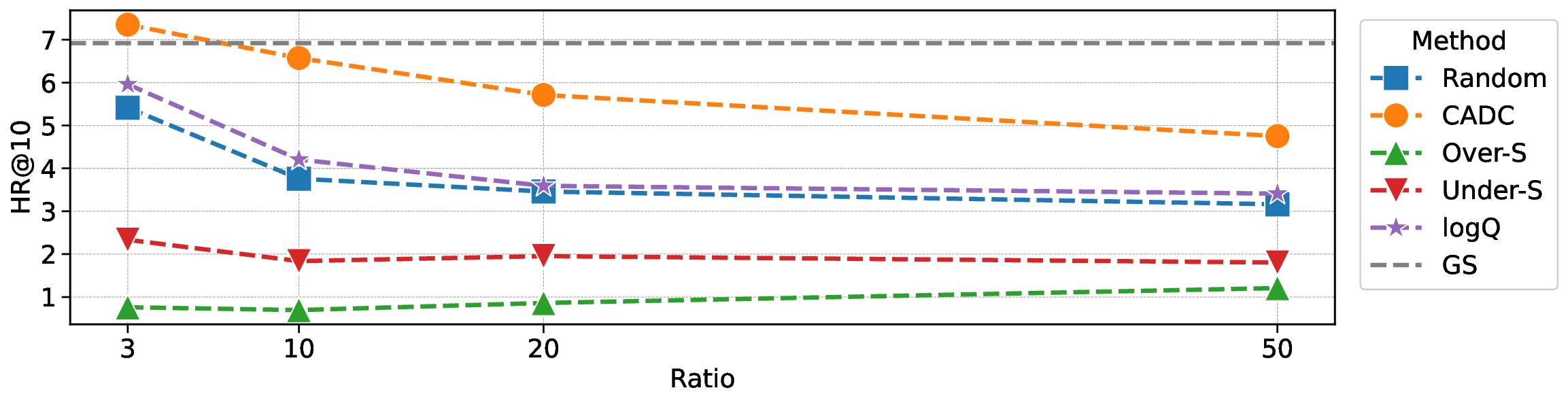}
\caption{Impact of Data Filtering Ratios on HR@10 for CADC. This plot demonstrates how the performance of CADC is influenced by varying levels of data reduction.}
\label{fig:sa_hr10}
\end{figure}

\begin{table}
    \centering
    \begin{tabular}{c|cc}
        \hline
        Method & HR@10 & NDCG@10\\
        \hline
        \textit{Hybrid} & 6.03 (12.9\%) & 3.08 (9.1\%)\\
        \textit{Init} & 6.23 (10.0\%) & 3.12 (8.0\%)\\
        \textit{Init-Frz} & 6.57 (5.1\%) & \textbf{3.28 (3.2\%)}\\
        \textit{Linear} & 6.56 (5.2\%) & 3.19 (5.9\%)\\
        \textit{MLP} & \textbf{6.59 (4.8\%)} & 3.22 (5.0\%)\\
        \hline
        GS & 6.92 & 3.39\\
        \hline
    \end{tabular}
    \caption{Comparative performance of CADC embedding integration techniques on MovieLens 1M, measured by HR@10 and NDCG@10. Percentages reflect deviation from the Gold Standard (GS).}
    \label{tab:emb_int}
    \vspace{-8mm}
\end{table}

The empirical results, displayed in Table \ref{tab:emb_int}, indicate that \textit{Init-Frz} not only simplifies computational demands with the fewest trainable parameters but also achieves superior outcomes compared to the alternatives. Specifically, it records the highest performance in NDCG@10 and closely competes with the more computationally demanding \textit{MLP} and \textit{Linear} methods in HR@10. Notably, allowing embeddings to be updated in the \textit{Init} method reduces accuracy, likely due to the noisy nature of TTNN gradients at the start of training and the limited data scope, which can compromise the integrity of pre-trained embeddings.

\section{Related Works}
\textit{Sampling Interaction Data.} Data Sampling is crucial in recommendation systems for extracting hard negative samples and assessing algorithms. It's been employed through methods such as random sampling, leveraging the underlying graph structure \cite{mittal2021eclare, ying2018graph}, and specific techniques such as similarity search \cite{jain2019slice} and stratified sampling \cite{chen2017sampling}. Sampling is also crucial in assessing recommendation algorithms \cite{canamares2020target, krichene2020sampled}. Additionally, it's also useful for createing smaller subsets of large datasets for purposes such as quick testing and algorithmic comparisons \cite{sachdeva2022sampling}.

\textit{Coreset Selection.} Coreset selection identifies data subsets that represent the full dataset's quality. Methods include score-based approaches, which select data based on criteria like forgetting frequency \cite{toneva2018empirical}, loss value \cite{kawaguchi2020ordered, jiang2019accelerating}, and prediction uncertainty \cite{coleman2019selection}, and gradient-based approaches, which estimate the dataset's gradient\cite{mirzasoleiman2020coresets, pooladzandi2022adaptive, yang2023towards}. These model-specific methods require computation at each data point, making them impractical for large datasets due to high computational demands. Our approach simplifies this by training a very basic computational model once, irrespective of the DLRM and content features to be used. This one-time training embeds collaborative information efficiently, circumventing the computational challenges of traditional coreset selection methods and facilitating scalable training for large datasets.

\textit{Data Distillation.} Data distillation synthesizes compact data summaries, primarily in continuous domains like images. These techniques distill the essential knowledge of an entire dataset into a significantly smaller, synthetic summary \cite{zhao2021dataset, nguyen2021dataset}. However, these techniques have predominantly focused on continuous data like images, a recent approach extended these methods to synthesize fake graphs, assuming pre-existing node representations, which limits their applicability to recommendation data \cite{jin2021graph}. Sachdeva et al. \cite{sachdeva2022sampling} adapt data distillation for collaborative filtering by generating high-fidelity, compressed data summaries specifically for use with infinitely-wide autoencoders ($\infty$-AE). This method is designed exclusively for $\infty$-AE applications in collaborative filtering and does not integrate content-based features. In contrast, our proposed method is developed for DLRMs like TTNN, incorporating both user-item interactions and content-based features into the training process.

\section{Conclusion}
This study introduces CADC, a pioneering approach designed to efficiently train DLRMs on large-scale datasets without substantially impacting model accuracy. Our findings demonstrate that by employing pre-trained embeddings that encapsulate comprehensive interaction data, CADC can significantly reduce the volume of data needed for training while preserving the collaborative information essential for maintaining high prediction quality. Tested across datasets like MovieLens 1M, MovieLens 10M, and Epinions, CADC not only outperforms traditional training methods in terms of efficiency and scalability but also maintains a high level of accuracy, closely approximating the performance of models trained on full datasets. Specifically, CADC has proven to mitigate the impact of data reduction on model performance, reducing training times dramatically without corresponding losses in effectiveness. Our research contributes to the broader field of recommendation systems by providing a scalable solution that addresses the twin challenges of maintaining high data throughput and model accuracy in the face of exponentially growing data sizes. It opens new avenues for future research, particularly in exploring more complex models and integration techniques that could further enhance the efficiency and effectiveness of recommendation systems.

\bibliographystyle{unsrt}
\bibliography{sample-base}

\end{document}